# Breakdown of continuum mechanics for nanometer wavelength rippling of graphene

Levente Tapasztó[1*], Traian Dumitrică[2], Sung Jin Kim[3], Péter Nemes-Incze[1], Chanyong Hwang[3] and László P. Biró[1]

1. Institute for Technical Physics and Materials Science, Research Centre for Natural Sciences H-1525 Budapest, Hungary. *Email: tapaszto.levente@ttk.mta.hu
2. Department of Mechanical Engineering, University of Minnesota, Minneapolis, Minnesota, 55455, USA
3. Center for Nano-imaging Technology, Korea Research Institute of Standards and Science, 267 Gajeong-Ro, Yuseong, Daejon 305-340, Korea

**Understanding how the mechanical behavior of materials deviates at the nanoscale from the macroscopically established concepts is a key challenge of particular importance for graphene, given the complex interplay between its nanoscale morphology and electronic properties[1,2,3,4,5]. In this work, the (sub-) nanometer wavelength periodic rippling of suspended graphene nanomembranes has been realized by thermal strain-engineering and investigated using Scanning Tunneling Microscopy. This allows us to explore the rippling of a crystalline membrane with wavelengths comparable to its lattice constant. The observed nanorippling mode violates the predictions of the continuum model[6], and evidences the breakdown of the plate idealization[7] of the graphene monolayer. Nevertheless, microscopic simulations based on a quantum mechanical description of the chemical binding accurately describe the observed rippling mode and elucidate the origin of the continuum model breakdown. Spatially resolved tunneling spectroscopy measurements indicate a substantial influence of the nanoripples on the local electronic structure of graphene and reveal the formation of one-dimensional electronic superlattices.**

---

[*] Correspondence should be addressed to L.T.

The nanoscale landscape is expected to markedly influence the electronic properties of the ideally two-dimensional graphene membranes[8,9,10]. When the local curvature of the graphene sheet is of the nanometer scale the electronic structure is substantially modified through altering the π-orbital energy (σ-π rehybridization[7]) and modifying the nearest neighbor hoping integrals that can induce a local shift of the electrochemical potential[8,9] as well as give rise to large pseudo-magnetic fields[3,4]. In fact, random nanoscale wrinkles are intrinsic to graphene[11], inducing spatially disordered charge density fluctuations, which act as disorder source for the propagating charge carriers[12,13]. By contrast, periodic ripples of nanometer wavelength – if achieved – comprise a much richer physics, giving rise to electronic superlattices[9], which fundamentally alter the quasi-particle energy dispersion and can render the propagation velocity of the charge carriers highly anisotropic[14]. Furthermore, it has been predicted that periodic graphene ripples can be employed for opening a (pseudo-) band gap[3,9]; that could eliminate the lithographic patterning of graphene and thus the problem of edge-quality.

Engineering of micron-scale periodic ripples in suspended graphene has been experimentally demonstrated[6] with wavelengths ranging from a few hundreds of nanometers up to several microns. In the reported rippling regime the results were found to follow the predictions of the continuum plate mechanics quite accurately[6].

To exploit the nanomechanical characteristics of graphene it is of particular interest to explore down to which length-scale the classical membrane behavior of graphene persists, and how it is altered below that. Furthermore, as the influence of ripples on the electronic structure, as well as chemical reactivity[15] of graphene is expected to drastically increase with the local curvature[8] downscaling the ripple wavelength in the nanometer regime is also of high practical relevance. To implement the nanoscale periodic rippling of graphene, one can employ the strain engineering of nanometer-sized suspended graphene membranes, as the rippling wavelength should scale with the square root of the membrane dimensions[16].

In this work, we show that such suspended graphene nanomembranes naturally form during the chemical vapor deposition (CVD) growth of graphene on Cu (111) single crystals. The STM investigation of the as-grown samples revealed the presence of reconstructed Cu surface areas underneath the continuous graphene layer. These reconstructions of the Cu (111) surface consist of monolayer height Cu adatom clusters as well as trench-like vacancy islands (see Fig. 1.a). Similar reconstructions have been previously reported on single crystalline Cu and Ag surfaces[17,18]. For more details concerning the sample preparation and the origin of the surface reconstructions see Supplementary Information. The trench-like vacancy islands have well-defined widths of about 5nm while their length is typically in the range of 20 to 80 nm. The atomic resolution topographic STM images reveal the honeycomb lattice of graphene all over the sample surface including the adatom pads and nanotrenches, these latter providing ribbon-like areas of suspended graphene nanomembranes. As evidenced from high resolution STM measurements, these nanomembranes display highly regular periodic oscillations, with a wavelength of ∼0.7 nm and modulation of ∼0.1 nm superposed on the honeycomb atomic lattice of graphene (Fig. 1b and Fig. 2). The reported nanorippling mode (with $\lambda = 0.7 \pm 0.1$ nm, $A = 0.5 \pm 0.2$ Å) has been observed on all of the more than thirty trenches investigated. Periodic oscillations in the STM images of graphene can often be attributed to electron interference[19]. Due to their nature these interference patterns are expected to strongly depend on the bias voltage (energy) used for the imaging[20]. However, in our case the experimentally observed pattern was practically unchanged (apart from a slight variation of the amplitude) upon varying the bias voltage. This is a clear signature of its topographic (structural) origin. We could also exclude the possibility that the observed features are Moiré-patterns[2,18,] as the graphene sheet is not in direct contact with the substrate above the trenches (for direct evidence see Supplementary Information). We propose the following mechanism for the formation of the graphene nanoripples. During the graphene growth, at temperatures of about 1000°C[21], a massive migration of the surface Cu atoms takes place underneath the growing graphene[22]. This gives rise to the observed surface reconstructions including the nanotrenches. Upon cooling down the sample from growth (~1000°C) to room temperature the Cu substrate contracts, while the overgrown graphene layer tries to expand according to its negative thermal expansion coefficient[6]. Consequently, where graphene is in direct contact, i.e. pinned to the Cu substrate it is subjected to a strong compressive stress as also suggested by previous Raman investigations[23]. However, in those areas where the graphene is suspended above the nanotrenches it is allowed to relax some amount of the compressive strain through out-of-plane deformation (rippling). The amount of strain due to the cooling down to RT can be approximated by: $\varepsilon \approx \Delta T(\alpha_{Cu} - \alpha_{Gr})$, where ΔT is the temperature difference between 1000°C and ambient, while $\alpha_{Cu}$ = 16.6 x $10^{-6}$ $K^{-1}$ and $\alpha_{Gr}$ = -7 x $10^{-6}$ $K^{-1}$ are the thermal expansion coefficient for Cu and graphene, respectively. This formula estimates the emerging compressive strain in the as-grown graphene samples to ε = 2.2%. Here we note that this is only a rough estimate as the TEC of graphene is also expected to be temperature dependent[24]. The strain can also be experimentally estimated from measuring the surface length of the graphene ripples[23], yielding similar strain values of a few percent. This thermally induced strain is responsible for the formation of the observed nanoripples in the suspended graphene areas.

As the curvature of a crystalline membrane becomes commensurate with the lattice constant, the applicability of continuum mechanics becomes uncertain. To verify, whether the classical behavior of graphene observed for micron scale ripples[6] remains valid also at the nanoscale, we attempted to interpret the observed graphene nanoripples within the established framework of continuum mechanics. The periodic ripples developed in a compressed thin plate of thickness *t* suspended over a trench of width *L* is characterized by [16, 25]:

$$\lambda^4 = \frac{4\pi^2 \nu (tL)^2}{3(1-\nu^2)\varepsilon} \quad (1a)$$

$$A^2 = \left[\frac{16\varepsilon\nu}{3\pi^2(1-\nu^2)}\right]^{1/2} tL \quad (1b)$$

Here, λ and *A* are the wavelength and amplitude of the rippling mode developed, while ε is the compressive strain along the trench edges.

Although derived and validated at the macroscopic scale, the above formulas have been successfully applied to describe the rippling of suspended (few-layer) graphene sheets at microscopic wavelengths (λ>300 nm)[6]. Compelled to apply them to the monolayer, one immediately faces the ambiguity of defining the thickness *t* for the single layer of C atoms. One approach[26] is to assign *t* = 3.35 Å, i.e. the experimentally-measured interlayer spacing in graphite. Using parameter values of *L* = 5 nm (membrane width) and ε = 2%, matching our experiments, together with the in-plane Poisson's ratio (ν = 0.16), these formulas give a wavelength of $\lambda_{theor}$ = 4.2 nm and an amplitude of $A_{theor}$ = 2.6 Å. This is in striking discrepancy with our measurements ($\lambda_{exp}$ = 0.7 nm, $A_{exp}$ = 0.5 Å). Varying the strain between 0.5% and 5% to account for the experimental uncertainty does not significantly improve the agreement. Another approach is to base our modeling on a plate with effective thickness *t* = 0.81 Å, derived from atomistically-computed in-plane (26.6 eV/A$^2$) and bending (1.6 eV/atom) constants of the monolayer[7,27]. This second model predicts a very different rippling mode, of wavelength $\lambda_{theor}$ = 2 nm and amplitude $A_{theor}$ = 1.3 Å, which is still in obvious discrepancy with the experimental findings.

Besides the values of the measured individual quantities (λ, A) we have also investigated the validity of the continuum mechanical equations relating them. A particularly useful relation to verify this can be established by combining the equations 1a and 1b. This way one can eliminate the strain (which cannot be accurately measured) and end up with: $A\lambda/L = \sqrt{\frac{8v}{3(1-v^2)}}\,t$. When supplying the corresponding parameter values (*A* = 0.05 nm, $\lambda$ = 0.7 nm, *L* = 5 nm, ν = 0.16, *t* = 0.335 nm) the above relation, clearly fails by an order of magnitude even for *t* = 0.81 Å.

The inability of the phenomenological continuum model to describe the peculiar sub-nanometer rippling observed in our experiments, prompted us to perform accurate microscopic-level simulations. We studied the rippling of freestanding graphene nanoribbons with fixed armchair edges by means of conjugate gradient relaxation simulations and a density functional-based tight binding interatomic potential[7,27], which describes explicitly the σ and π C-C bonding of graphene. To closely mimic the boundary conditions with a reduced number of atoms, the simulation domain containing 600 atoms was placed under periodic boundary conditions along the z dimension, which corresponds to the nanotrench long dimension. The trench width dimension was modeled explicitly as the distance between the two strained dimer lines located at the edges measures $L$=5.7 nm. For further technical details of the simulation method see Supplementary Information. In contrast to the continuum model, we found that the microscopic approach can regain sub-nanometer rippling and thus confirms our interpretation of the STM observations. Indeed, Fig. 3a displays the periodic rippling exhibited by a graphene ribbon subjected to 5% edge biaxial compression. This metastable rippling mode presents an energy advantage of 62 meV/atom over the compressed planar morphology. Its wavelength of 0.8 nm and amplitude of 0.5 Å (Fig. 3b) are both in excellent agreement with our experimental results.

The striking discrepancy with the $t$ = 3.35 Å plate-based model predictions is a manifestation of the breakdown of the plate phenomenology[7]. In the monolayer, the resistance to bending involves the π-orbital misalignment between adjacent pairs of C atoms, and can be completely decoupled from its in-plane σ bond deformations. Thus, an atomic membrane with such stiff in-plane bonding as the σ bonds of graphene can ripple at sub-nanometer scale, which implies almost no resistance against out of plane deformations. This behavior is in sharp contrast with the phenomenology captured by the classical continuum plate, where the bending of a plate always induces the in-plane stretching and compression on the opposite sides of a neutral curved surface. Nevertheless, the malfunctioning of the continuum model goes beyond the issue of selecting an appropriate plate thickness. The failure of the $t$ = 0.81 Å second plate model, which formally gives a bending constant in agreement with the above microscopic model, can be understood based on the bond length analysis displayed in Fig. 3c. Recall that in both the equilibrium flat and inextensional bending case, the C-C bonds measure 1.42 Å. In the relaxation through sub-nanometer rippling, the C-C bonds still store a significant amount (~3.5%) of compressive strain. Therefore, the breakdown of the plate phenomenology allows the violation of the inextensible deformation assumption, which is key for deriving equations (1)[16]. Of course, the experimental mechanical constrains set by the nanotrenches do not prohibit the onset of rippling modes in which bending would be accomplished in an approximately inextensible manner.

Since the atomistic mechanisms allow a much easier out-of-plane deformation of graphene at the nanoscale as compared to classical plates, this ultrasoft bending behavior might also be at the origin of the ultrastrong adhesion of the monolayers to different substrates[28], by allowing a substantially better conformation to the substrate irregularities. In comparison, for multilayers, the interlayer coupling reestablishes the coupling between bending and in-plane deformations[7], abruptly increasing the flexural rigidity[29].

We have also performed spatially resolved tunneling spectroscopy measurements[2,10,12] to verify how the nanoscale structural ripples affect the local electronic properties of graphene[8,9,30]. In this measurement mode individual tunneling I-V characteristics are acquired with the tip fixed over a specific location of the sample and the sample voltage ramped within a given window while measuring the corresponding tunnel current (for further details see Supplementary Information). Figure 4 shows the differential tunneling conductivity (~LDOS) map obtained by plotting the numerical derivative of the individual current-voltage characteristics at each location for a sample bias of $U_{bias}$ = 48 mV. The spectroscopy map reveals the periodic modulation of the local density of electronic states (LDOS) on the nanorippled region. This evidences that nanoscale structural ripples substantially affect the local electronic structure of graphene giving rise to one-dimensional electronic superlattices.

The ability of graphene to ripple down to sub-nanometer wavelengths can be exploited in strain-engineering graphene based nanoelectronic and nanoelectromechanical devices beyond the boundaries set by continuum mechanics.

**Acknowledgements**

The experimental work has been conducted within the framework of the Korean-Hungarian Joint Laboratory for Nanosciences through the Converging Research Center Program (2010K000980). L.T. acknowledges OTKA grant PD 84244. P.N. and L.P.B. acknowledge OTKA grant K 101599. T.D. acknowledges NSF CAREER Grant CMMI-0747684. L.T. is grateful to the Alexander von Humboldt Foundation.

**Author contributions**

L.T. conceived and designed the experiments. L.T. and P.N. performed the STM experiments. S.J.H and C.H. performed the growth experiments. T.D. provided the simulations results. L.T., T.D. and L.P.B analyzed the data. L.T. and T.D. wrote the paper. All the authors discussed the results and commented on the manuscript.

**Figures and captions**

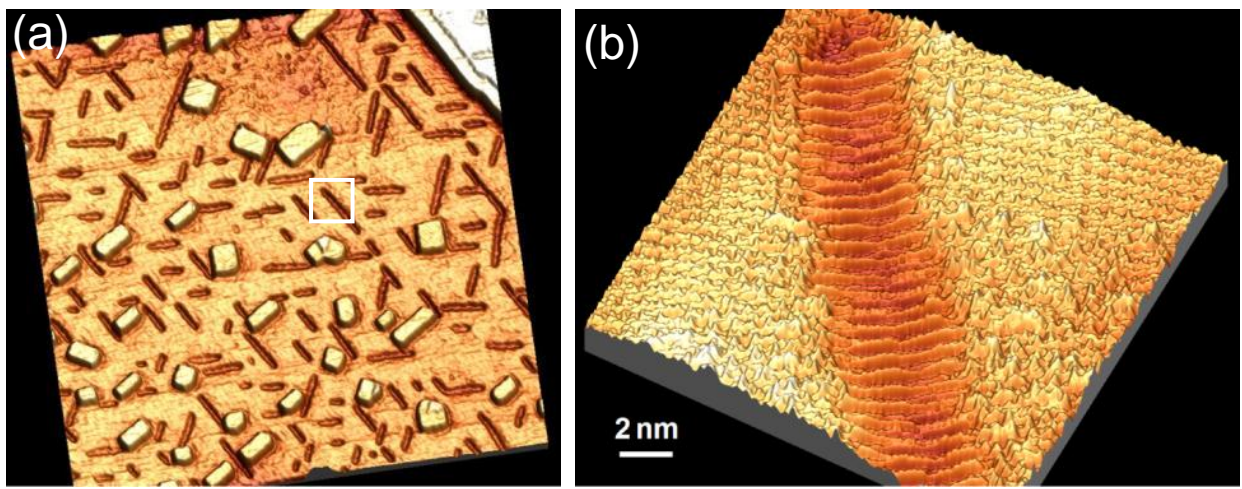

Figure 1. **3D STM images of nanotrenches and graphene nanoripples**. a) STM image (300x300 nm) of a reconstructed Cu(111) surface continuously covered by graphene. The rectangular protrusions are single atom height Cu adatom clusters, while the trench-like features correspond to vacancy islands with well-defined widths of 5 nm and oriented along three particular directions. b) High resolution STM image of a nanotrench revealing the nanoscale periodic rippling of the graphene membrane suspended over the trench.

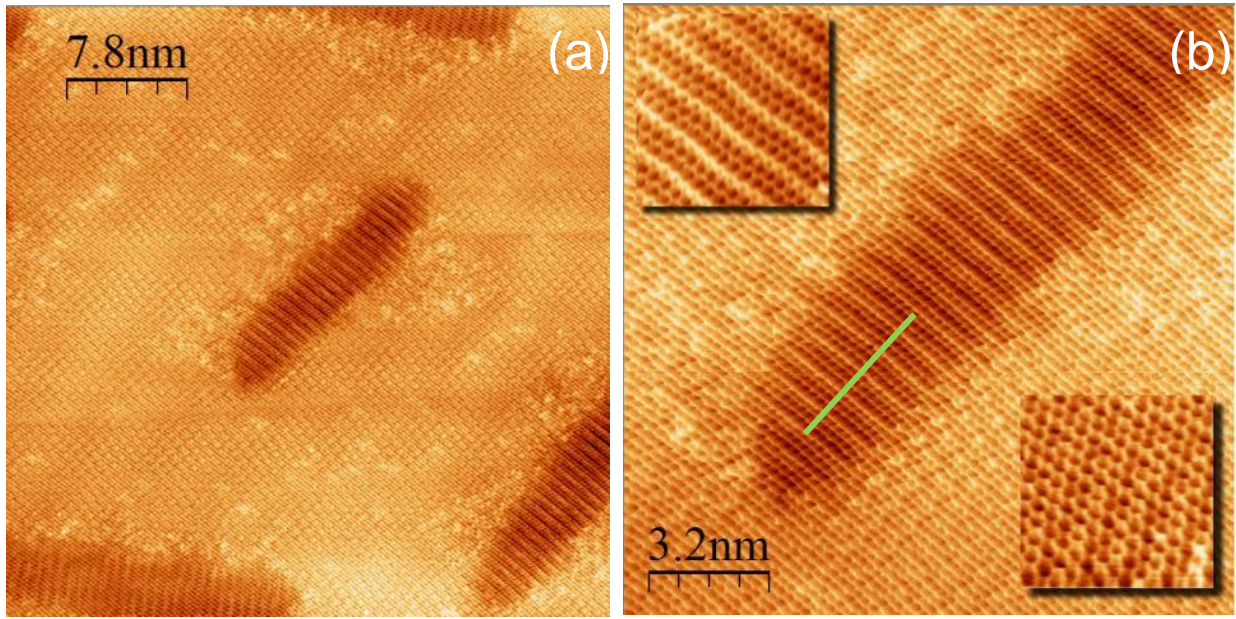

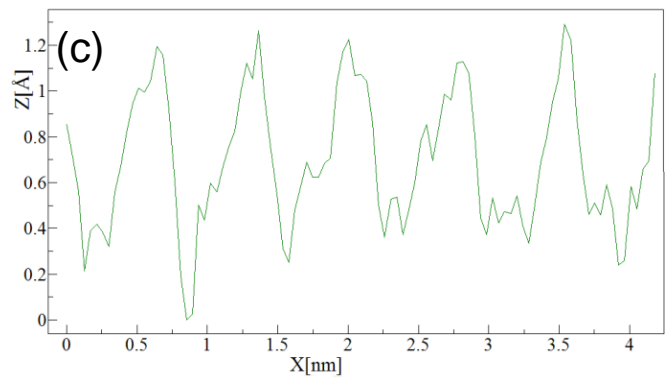

Figure 2 **Atomic resolution STM images of graphene nanoripples** a) STM image showing several nanotrenches of different orientations, all displaying graphene nanoripples over the trenches with the ripple crests always perpendicular to the trench edges. On the flat regions between the trenches a Moiré-pattern can be observed. b) Atomic resolution STM image of a nanotrench displaying sub-nanometer graphene ripples. The magnified insets display the honeycomb graphene lattice both over the flat substrate (bottom right) and the rippled region over the trenches (top left). c) Cross section of the graphene ripples displaying a wavelength of $\lambda$ = 0.7 nm, and amplitude of A = 0.05 nm. The additional corrugation of the line can be attributed to the positions of C atoms.

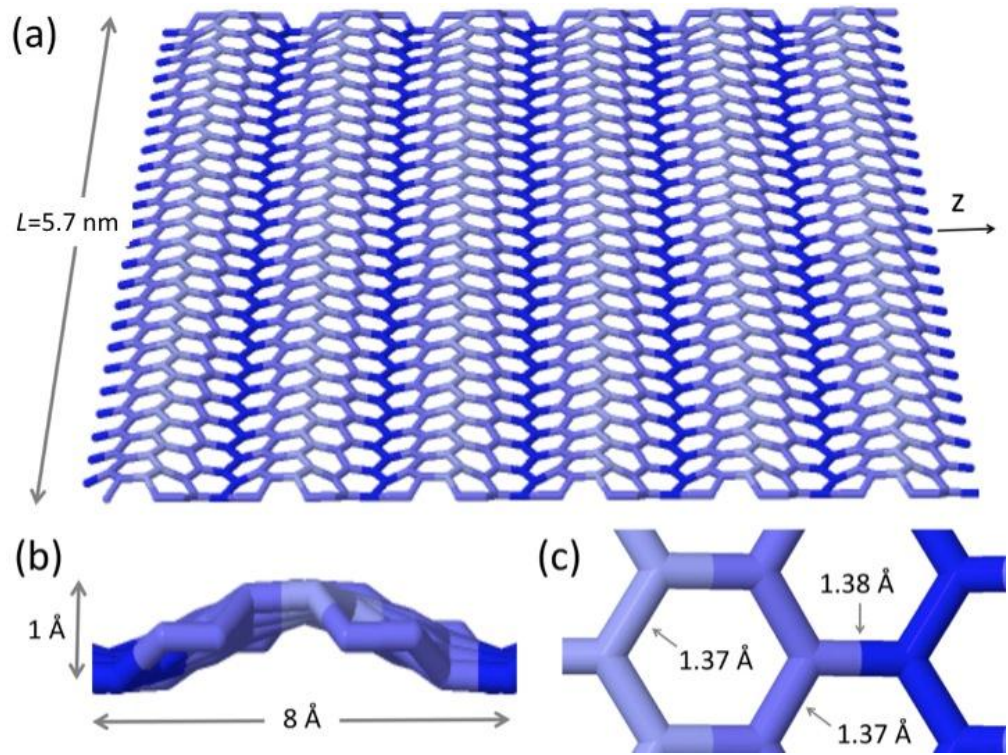

Figure 3. **Atomic-scale simulations of graphene nanoripples**. a) Simulation of a 5.7 nm wide infinitely-long graphene nanoribbon subjected to 5% in-plane compression, revealing a periodic rippling morphology of 0.8 nm wavelength and 1 Å modulation (b). c) Details of the C-C bond lengths; the color scale indicates the height profile with light (dark) blue atoms at 0.5 (-0.5) Å heights.

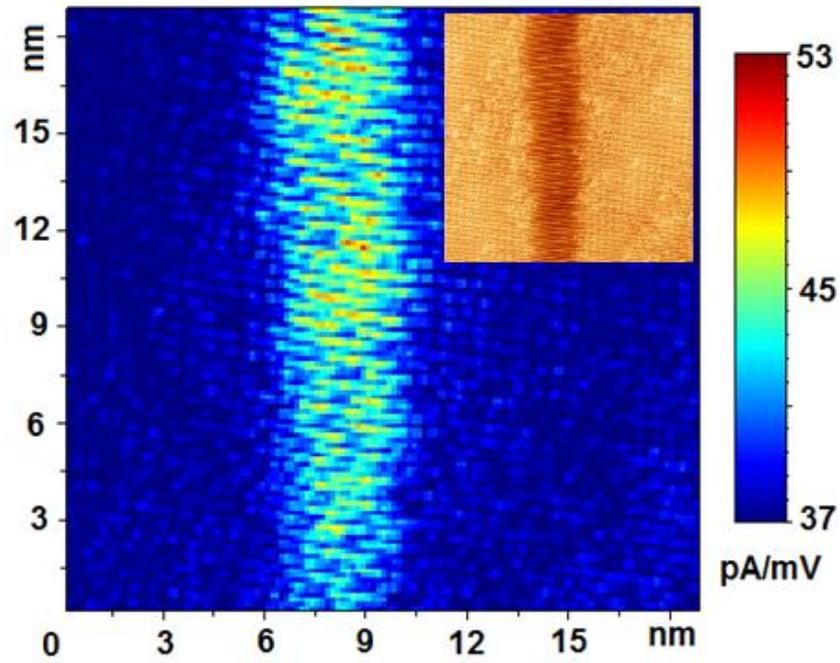

Figure 4. **Local electronic density of states map on graphene nanoripples**. Spatially resolved differential tunneling conductivity map (plotted at $U_{bias}$ = 48 mV) displaying the periodic modulation of the local density of states on the graphene nanoripples. The inset shows the corresponding topographic image.